\documentclass{aa}
\usepackage{psfig}
\newcommand{\NH}{$N_{\rm H}$}

\newcommand{\ergscm}{erg\,s$^{-1}$\,cm$^{-2}$}

\begin{document}

    \thesaurus{03       
              (04.19.1;  
               13.25.3   
               13.25.2   
               13.25.5   
               11.01.2   
               11.03.1   
             )}

   \title{Identification of a complete sample of northern ROSAT
All-Sky Survey X-ray sources} \subtitle{IV. Statistical analysis}

%
\author{J. Krautter \inst{1} \and F.-J.~Zickgraf
\thanks{Present address: Hamburger Sternwarte, 
Gojenbergsweg 112, D-21029 Hamburg, Germany}
\inst{2} \and
I. Appenzeller \inst{1} \and I. Thiering
\thanks{Present address: Max-Planck-Institut f\"ur 
Astronomie, K\"onigstuhl 17, D-69117 Heidelberg, Germany}  
\inst{1,}\inst{3}
\and W. Voges \inst{4} \and  C. Chavarria \inst{5}
\and R. Kneer \inst{1} \and R. Mujica \inst{6}
\and M.W. Pakull \inst{2} \and A. Serrano \inst{6}
\and B. Ziegler \inst{7}
}  

\institute{Landessternwarte K\"onigstuhl, D-69117 Heidelberg, Germany
\and Observatoire Astronomique de Strasbourg, 11, rue de l'Universit\'e,
F-67000 Strasbourg, France
\and Max-\-Planck-\-Institut f\"ur 
Astronomie, K\"onigstuhl 17, D-69117 Heidelberg, Germany
\and Max--Planck--Institut f\"ur extraterrestrische Physik,
 Giessenbachstrasse, Postfach 1603, D-85740 Garching, Germany
\and Instituto de Astronomia-UNAM, Apdo. Postal 70-264, 04510 M\'exico 
D.F., Mexico
\and Instituto Nacional de Astrofisica, Optica y Electronica (INAOE), 
A. Postalo 51 y 216 Z.P., 72000 Puebla, M\'exico
\and Dept. of Physics, Univ. of Durham, South Rd, Durham DH1 3LE, United 
Kingdom }

\offprints{J. Krautter} 
\date{Received date; accepted date}

\maketitle

\begin{abstract}
We present the statistical evaluation of a count rate and area limited
complete sample of the ROSAT All-Sky Survey
(RASS) comprising 674 sources. The RASS
sources are located in six study areas outside the galactic plane
($|b^{II}| \ga 20^{\circ}$) and north of $\delta = -9^{\circ}$.  The
total sample contains 274 (40.7\%) stars, 26 (3.9\%) galaxies, 284
(42.1\%) AGN, 78 (11.6\%) clusters of galaxies and 12 (1.8\%)
unidentified sources.  
These percentages vary considerably 
between the individual
study areas due to different mean hydrogen column
densities \NH . For the accuracy of the RASS positions, i.e., the
distance between optical source and X-ray position, a 90\% error
circle of about 30\arcsec\ has been found.  Hardness and
X-ray-to-optical flux ratios show in part systematic differences between the
different object classes. However, since the 
corresponding parameter spaces overlap significantly for the 
different classes, an unambiguous identification on the basis of the
X-ray data alone is not possible.
The majority of the AGN in our sample
are found at rather low redshifts with the median value of the whole
sample being $z_{\rm med}=0.24$. The median values for the individual
fields vary between $z_{\rm med}=0.15$ and $z_{\rm med}=0.36$.
The slope of the $\log\,N\,-\,\log\,S$ distributions for our AGN ($-1.44
\pm0.09$) is in good agreement with the euclidian slope 
of $-1.5$. The BL
Lacs (with a slope of $-0.72\pm0.13$) exhibit a much flatter distribution.

\end{abstract}

\section{Introduction}
\label{intro}
The ROSAT All-Sky Survey ({\sl RASS}) was the first all-sky survey in
the soft X-ray range (0.07 keV to 2.4 keV) with an imaging telescope
(Aschenbach \cite{Aschenbach88}). It started on July 30, 1990, two
months after the launch of the ROSAT X-ray satellite (Tr\"umper
\cite{Trumper83}) on June 1, 1990. Using the Position Sensitive
Proportional Counter ({\sl PSPC}) ROSAT scanned the sky during the All-Sky
Survey in great circles perpendicular to the direction of the sun, and
thus covered the whole sky in half a year. The RASS allowed for the
first time to study unbiased, {\sl spatially complete} samples of
X-ray sources. Previous knowledge of the composition of the 
X-ray sky at high galactic latitudes was mainly based on the
{\sl EINSTEIN Medium Sensitivity Survey} ({\sl EMSS}) (Stocke et al. 
\cite{Stockeetal91}) containing serendipitious sources found
in X-ray images of fields with otherwise scientifically interesting
objects.

On the basis of the X-ray data alone it is usually not possible to
determine the physical nature of the X-ray source. In order to do so,
optical follow-up observations have to be carried out. Since during
the RASS about 60\,000 sources were detected (Voges et
al. \cite{Vogesetal99}), it is obvious that a complete identification
of all sources is beyond any reasonable scope. In order to obtain
representative subsamples six study areas from the RASS were selected
for a complete optical identification. These study areas are located
north of $\delta = -9^{\circ}$, with $|b^{II}| \ga 20^{\circ}$
including one region near the North Galactic pole (NGP) and one near
the North Ecliptic Pole (NEP).

A detailed description of the fields, the selection criteria, the
optical observations and the criteria for the optical identification
bave been extensively discussed in Zickgraf et al. (\cite{Zickgrafetal97a}),
hereafter Paper\,II.  As described in Paper\,II, in order
to reduce the sample to a manageable but still statistically
significant size, we chose minimum count rates of 0.03 cts s$^{-1}$
for five of our areas and 0.01 cts s$^{-1}$ for one area.  A catalogue
of the complete optical identifications of our flux limited sample is
presented in Appenzeller et al. (\cite{Appenzelleretal98}), 
hereafter Paper\,III.
On the basis of this paper we shall give in the following a detailed
statistical analysis of the final results from our identification
programme.

\section{Basic Statistics} 
\label{Basic Statistics} 

Some basic statistical information on the optical identifications in 
our six study
areas is given in Table 1. The field designations 
are essentially those defined in Table 1 of Paper\,II. 
However, for study areas IV and V
we have used a slightly modified field definitions: Study area IV of
this paper corresponds to IVac of Paper\,II, and study area V to Va of
Paper\,II.  We note, that the designations used here
correspond to those already used in Paper\,III.

The first row of Table 1 gives the number of sources in the individual study
areas which fulfill our flux criteria mentioned above. In the second
row the number of sources is given for which the identification
exclusively rests on a comparison with the SIMBAD or NED data bases.
The third row gives
the number of sources observed by ourselves (91.8\% of all sources).

\begin{table}[ht]
\caption[]{Basic statistical information on the identifications.
The class ``stars'' includes
normal stars, Ke, Me stars, CVs, binaries, and white dwarfs.  }
\begin{tabular}{lrrrrrrr}
\noalign{\smallskip} \hline \noalign{\smallskip}
\multicolumn{1}{c}{study area}
&\multicolumn{1}{c}{I}&\multicolumn{1}{c}{II}&\multicolumn{1}{c}{III}
&\multicolumn{1}{c}{IV}&\multicolumn{1}{c}{V}&\multicolumn{1}{c}{VI}
&\multicolumn{1}{c}{total} \\ 
\noalign{\smallskip} \hline \noalign{\smallskip} 
No. of sources: & 100 & 122 & 124 & 110 & 125 & 93 & 674\\ 
SIM/NED id.: & 6 & 20 & 12 & 8 & 6 & 3 & 55 \\ 
Observed: & 94 & 102 & 112 & 102 & 119 & 90 & 619 \\ 
\noalign{\smallskip} \hline \noalign{\smallskip} 
Identified: &100 & 120 & 120 & 109 & 123 & 90 & 662 \\
stars & 63 & 52 & 43 & 13 & 43 & 34 & 248 \\ 
galaxies & 2 & 6 & 9 & 4 & 0 & 2 & 23 \\ 
AGN & 19 & 44 & 48 & 68 & 55 & 38 & 272 \\ 
clusters of gal. & 7 & 8 & 11 & 14 & 22 & 10 & 72 \\ 
multiple sources & 9 & 10 & 9 & 10 & 3 & 6 & 47 \\ 
no id.  & 0 & 2 & 4 & 1 & 2 & 3 & 12 \\
\noalign{\smallskip} \hline
\end{tabular}
\label{idents}
\end{table}

In the lower part of the table the number of identified sources 
for different object classes as well as 
the number
of unidentified sources is given. Only for a total of 12 sources (1.8\%) 
no identification could be found. This fraction is lower by
more than a factor of two than the 3.9\% unidentified sources in the
EMSS (Stocke et al. \cite{Stockeetal91}).  As already mentioned in
Paper\,III, there are 21 sources (3.1\%)
for which we have a likely, but uncertain identification
('identification quality index' Q=3 in Table 1 of Paper\,III). A
discussion of these sources can be found in the comments on the
individual sources in Paper\,III. In the following we shall 
not distinguish these cases from the reliable identifications. 

For 47 X-ray sources (7.0\%) ('mult. sources' in Table 1) two possible
optical counterparts were found (cf. the discussion in Paper\,II). In
Paper\,III both sources are listed in the order of likelihood or suspected 
relative contribution to the total X-ray flux. Table 2 lists the
number of 'pairs' of object classes for these multiple sources:

\begin{table}[ht]
\caption[]{Distribution of multiple sources according to object classes}.
\begin{tabular}{lr}
\noalign{\smallskip} \hline \noalign{\smallskip}
star - star:        & 19 \\
star - AGN:         & 11 \\
star - gal:         &  1 \\
star - galaxy cl:   &  6 \\
AGN - AGN:          &  2 \\
AGN - gal:          &  2 \\
AGN - galaxy cl:    &  2 \\
gal - galaxy cl:    &  4 \\
\noalign{\smallskip} \hline
\end{tabular}
\label{idents}
\end{table}

Of the 47 multiple sources 37 (78\%) contain at least one star.
For our statistical purposes we consider as the optical
identification the more likely source which is listed first in
column 12 of Paper\,III. (The same procedure was applied by Stocke et al. 
\cite{Stockeetal91} for the EMSS.)

\subsection{Distribution of sources} 
\label{Distribution of sources} 

With the inclusion of the more plausible parts of the multiple sources 
we obtain the following distribution of all
identified sources for the complete sample: stars: 274 (40.7\%), galaxies:
26 (3.9\%), AGN: 284 (42.1\%) and clusters of galaxies: 78 (11.6\%). 
Table 3 lists this distribution along with the distribution in the individual
fields and a detailed subdivision of the object classes.

\begin{table}[ht]
\caption[]{Number distribution of optical counterparts for the
different study areas.
The abbreviations mean: ES: hot emission line star
(including cataclysmic variables), 
S: Seyfert galaxy, Nl1: Narrow line Seyfert 1 galaxy,
AGN uncl.: Active galactic nucleus for
which a more accurate classification was not possible from our spectra, 
int. gal.'' interacting galaxy.}
\begin{tabular}{lrrrrrrr}
\noalign{\smallskip} \hline \noalign{\smallskip}
\multicolumn{1}{c}{study area}
&\multicolumn{1}{c}{I}&\multicolumn{1}{c}{II}&\multicolumn{1}{c}{III}
&\multicolumn{1}{c}{IV}&\multicolumn{1}{c}{V}&\multicolumn{1}{c}{VI}
&\multicolumn{1}{c}{total} \\ 
\noalign{\smallskip} \hline \noalign{\smallskip} 
No. of sources:  & 100 & 122 & 124 & 110 & 125 & 93 & 674 \\
Stars:           & 70 & 60 & 48 & 15 & 44 & 37 & 274 \\
~~A stars:       &  2 &  0 &  1 &  0 &  0 &  0 &   3 \\
~~F stars:       & 13 & 14 &  8 &  2 & 12 &  4 &  53 \\
~~G stars:       & 15 & 16 &  8 &  3 &  2 & 10 &  54 \\
~~K stars:       & 11 &  7 & 11 &  4 & 13 &  8 &  54 \\
~~M stars:       &  0 &  1 &  1 &  1 &  1 &  1 &   5 \\
~~Ke stars:      &  9 &  9 &  7 &  0 &  5 &  5 &  35 \\
~~Me stars:      & 17 &  7 & 10 &  5 &  8 &  6 &  53 \\
~~ES:            &  1 &  6 &  2 &  0 &  2 &  3 &  14 \\
~~WD:            &  2 &  0 &  0 &  0 &  1 &  0 &   3 \\
Act. gal. nuc.:  & 20 & 45 & 50 & 74 & 56 & 39 & 284 \\
~~S1:            & 11 & 22 & 32 & 43 & 20 & 18 & 146 \\
~~S1.5:          &  1 &  0 &  0 &  3 &  2 &  4 &   9 \\
~~S2:            &  0 &  2 &  3 &  4 &  3 &  4 &  16 \\
~~LINER:         &  0 &  2 &  0 &  1 &  0 &  0 &   3 \\
~~NL1:           &  1 &  1 &  0 &  0 &  0 &  1 &   3 \\
~~QSO:           &  3 &  8 &  9 & 18 & 24 &  8 &  70 \\
~~BL Lac:        &  3 &  9 &  4 &  4 &  7 &  3 &  30 \\
~~AGN uncl.:     &  1 &  1 &  2 &  1 &  1 &  1 &   7 \\
Galaxies         &  3 &  6 & 11 &  4 &  0 &  2 &  26 \\ 
~~normal:        &  2 &  2 & 10 &  2 &  0 &  1 &  17 \\
~~int. gal.:     &  0 &  1 &  0 &  1 &  0 &  0 &   2 \\
~~radio gal.:    &  1 &  3 &  1 &  1 &  0 &  1 &   7 \\
Galaxy clusters: &  7 &  9 & 11 & 16 & 23 & 12 &  78 \\ 
\noalign{\smallskip} \hline
\end{tabular}
\label{idents}
\end{table}

Nearly the same fraction of stars and AGN are found in our 
RASS sample (40.7\% and 42.1\%, respectively). Together they represent
nearly 83\% of all sources. This differs
significantly from the distribution of sources in the EMSS (Stocke et
al. \cite{Stockeetal91}).  The RASS sample shows a much higher
fraction of stellar counterparts than the EMSS (40.7\% vs. 25.8\%),
but a lower fraction of AGN (42.7\% vs. 54.5\%). The fraction of
clusters of galaxies is with 11.6\% (RASS) and 12.2\% (EMSS) about the
same in both samples.  The same applies for galaxies with 3.9\% (RASS)
and 2.7\% (EMSS) which do not show significant differences either.

More than half of all Active Galactic Nuclei were found to be
Seyfert 1 galaxies. Alltogether the Seyfert galaxies form
62.3\% of the AGN counterparts. About 25\% are QSO and the fraction
of BL Lac objects amounts to 10.6\% This fraction corresponds to
4.5\% of the whole sample which is nearly the same fraction
as found for the EMSS. However, among the AGN the fraction of the BL Lac 
objects is with 7.8\% somewhat lower in the EMSS. 

\subsection{Stellar counterparts } 
\label{Stellar counterparts}

Among the stellar counterparts K stars (including Ke stars) 
occur most frequently (32.5\%). F, G and M/Me stars have about the same
frequency (19.3, 19.7, and 21.2\%, respectively). About 39\% of the K
star and more than 90\% of the M star counterparts display emission
lines.  This obviously reflects the fact that emission lines in the optical
spectrum indicate strong stellar activity which tends to give rise to
X-ray emission.  The frequency of Ke stars may have been 
underestimated, since at our low spectral resolution weak emission
lines could not always be detected.

In order to check whether some of our stellar identifications could be
due to a random positional coincidence with field stars, we calculated
the number of stars $n_{\rm err}$ expected for the total area of our 674
error circles (0.147 deg$^{2}$). For our calculations we used the
luminosity function by spectral type presented in Table 4-7 by 
Mihalas \& Binney 
(\cite{MihalasBinney81}). 
For the scale heights in $z$ direction we used the
numbers given in Table 4-16 by Mihalas and Binney. The numbers for
$n_{\rm err}$ were calculated for Galactic Latitude $b^{\rm II}$ of 45
degrees. Interstellar extinction was not taken into account.

For both F and M stars random positional
coincidences do not play any significant role. For F stars brighter
than our detection limit of $V \sim$ 14 mag we expect 0.8 stars in the
total area of our error circles, for M stars brighter than the limit
of $V$ = 15 mag we expect 0.3 stars, respectively.

The situation is different, however, for G and K stars. Here we expect
for stars down to a brightness limit of $V$ = 15 mag some 7.6 and 6.6
stars, respectively, in the total area of our error circles.  For a
brightness limit of 13 we expect for the G stars still 3.4 stars in
the error circles, whereas for the K star the expectation value of 0.8
stars drops below one. The expectation value of one star per error
circle area is reached for G stars at $V \sim$ 12 mag.

These results show that the number of dwarf stars randomly present in
the total area of our error circles should be of the order of $\sim$15
sources. However, the actual number of G stars fainter than 12th
magnitude and K stars fainter than 13th magnitude incorrectly
identified as X-ray sources due to a random positional coincidence,
should be much lower. Firstly, as all G and K stars emit X-rays at
some level, it is unlikely that all the stars in the error circles are
unrelated to the observed X-ray flux. Secondly, as most error circles
contain another plausible optical counterpart (such as an AGN), the
majority of the randomly present dwarf stars are expected to appear as
the less probable component in multiple sources. (There are nine
multiple sources, where we considered a faint G or K star to be the
less likely counterpart). Thirdly, a source with, for instance, a
bright AGN and a faint dwarf star normally has not been classified as
multiple source.  Therefore, even with most pessimistic assumptions a
random positional coincidence should not have resulted in more than
about two to three G and the same number of K stars incorrectly
identified as optical counterparts of the observed X-ray
sources. These numbers are significantly smaller than the $\sqrt{n}$
statistical uncertainties of our results for these stars and,
therefore, do not affect our statistical conclusions.

Our resolution was, unfortunately, not sufficient, to detect the lithium
$\lambda$6707 absorption line which would be an indicator for a
relatively young age of the star.  In a follow-up study of the present
identification project, Zickgraf et al.  \cite{Zickgrafetal98} found
among 35 K and M stars of field I nine stars which display significant Li
$\lambda$6707 absorption. First results of a detailed study of field
VI by Zickgraf et al. (in preparation) indicate that here the fraction
of Li stars among the K stars is lower than in field I.

We note that three A stars (two A2, one A5) were found in our
sample. This is unexpected, since, according to theoretical
predictions, early to mid A stars should not show any X-ray emission
because of the missing convection zone (e.g. Haisch \& Schmitt
\cite{Haisch96}).  In most cases, where X-ray emission seemed to be
present in A stars, it was found to originate from a cool companion.

\subsection{Comparison of individual fields}
\label{Individual fields}

As can be seen from Table 3, there are in part 
distinct differences between
the individual fields. 
While in field I the stellar counterparts
outnumber the AGN by a factor of 3.5, there are 5 times more AGN than
stars in field IV. These differences are of high statistical
significance.
The main reason for this differing ratios are
different mean hydrogen column densities \NH . In field I $<N_{\rm H}>
(11.5\cdot10^{20}$\,cm$^{-2})$ is more than seven times higher than
in field IV (1.6$\cdot10^{20}$\,cm$^{-2}$). The main
reason for this strongly differing hydrogen column density is 
the location at different galactic latitudes $b^{II}$. While
field I is at an intermediate galactic latitude $b^{II} \sim
-37^{\circ}$, field IV is close to the North Galactic Pole (mean
latitude $b^{II} \sim +84^{\circ})$.

In fields II, III, V, and VI the ratios n$_{stars}$/n$_{AGN}$ are
between 0.79 (field V) and 1.33 (field II). In fields III and VI about
the same numbers of stars and AGN have been identified. These four
fields have - with $<N_{\rm H}>$ values between 3.9$\cdot10^{20}$ and
5.5$\cdot10^{20}$\,cm$^{-2}$ - intermediate mean hydrogen column
densities. For these fields no correlation between the
ratio of stars and AGN with $<N_{\rm H}>$ could be found.

\subsection{Accuracy of the RASS positions} 
\label{Accuracy of the RASS positions} 

\begin{figure}
\psfig{figure=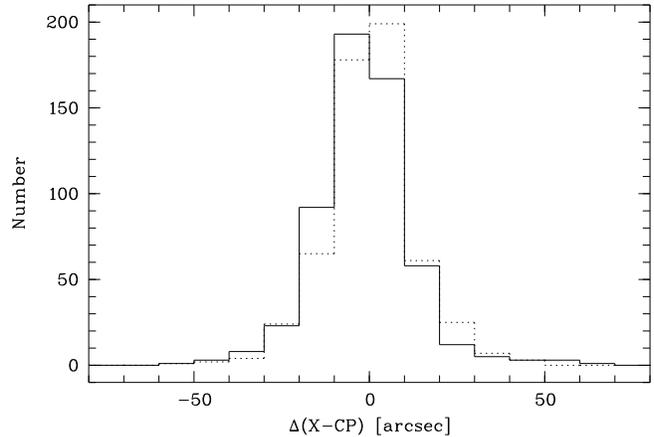,width=8.5cm,angle=-90,bbllx=52pt,bblly=28pt,bburx=543pt,bbury=741pt,clip=}
\caption[]{Histograms of the positional uncertainties. 
The differences in arcsec between X-ray
position ('X') and optical counterpart position ('CP') are 
plotted separately for
RA (solid line) and DEC (dashed line). 
}
\label{jk8432.f1}
\end{figure}

With our complete sample it is possible to derive final values for
the positional uncertainty of RASS X-ray sources, i.e., 
the mean distance of the
identified optical sources from the X-ray positions. We follow the
same procedure as outlined in section 5.2 of Paper\,II. Figure 
\ref{jk8432.f1} shows
the positional uncertainties in right ascension (solid line) and in
declination (dashed line) of our sources.  1\,$\sigma$, (i.e. the 67\%
error circle) corresponds to 13.5\arcsec\ for right ascension and 
12.8\arcsec\ for
declination, i.e. about the same for each coordinate. The sytematic offset
towards the east due to the motion of the satellite during read-out of
the data is 1.5\arcsec . The standard deviation found here is about
4\arcsec\ larger than in the preliminary data presented in Paper\,II.

For the total distance between the optical source and the X-ray
position we derive a 1\,$\sigma$ standard deviation of 17\arcsec\ and
a 90\% error circle of 30\arcsec . These numbers are significantly
smaller than those calculated by the SASS (Standard Analysis Software
System) where a 90\% error circle of 40-45\arcsec\ was derived (Voges
et al., in preparation) and also smaller than the 50\arcsec\ 90\%
confidence radii from the EMSS (Stocke et al.  \cite{Stockeetal91}).

Because of the smaller area of the error circle of the RASS as
compared to the EMSS, confusion of X-ray sources should be lower
in the RASS than in the EMSS. Stocke et al. (\cite{Stockeetal91})
found for their 
EMSS a source confusion level of the order of 2.5\% (21 out
of 835 sources). Since the sensitivity levels of the RASS and the EMSS
are not too different, we conclude that source confusion in our
flux limited RASS sample should be well below 1\%.

\section{X-ray, optical and radio quantities}
\label{quantities}

\subsection{Hardness ratios}
\label{Hardness ratios}

Tables 4 and 5 show mean hardness ratios of our samples. (For the
definition of the hardness ratios see e.g. Paper\,II). For the
calculation of the mean values only hardness ratios with internal
errors $\sigma \leq 0.5$ have been used. In addition, all hardness
ratios below -1.0 and above +1.0 were set to -1.0 and +1.0,
respectively. Values of $|HR1| \geq 1.0$ are purely artificial
without physical meaning. They arise from the fact that for very soft
or and for very hard sources the X-ray flux in the hard or soft range,
respectively, is below the detection limit. So fluctuations in the
background subtraction can give (physically meaningless) negative
values in these bands.

For the individual study areas mean hardness ratios have been
calculated for stars and AGN only. For the three other groups, galaxies,
cluster of galaxies, and sources with no optical identification (which
are included for completeness reasons) the number of objects in the
individual fields is too low to get any significant results.

As Tables 4 and 5 show, for the mean values of HR2, which subdivides
the hard range, there are only marginal differences between the
individual classes of objects. This applies to both the total sample
and to the individual fields. Even in the case of strongly differing
$<$HR1$>$ values are found (like for the AGN in fields I and IV, +0.83
vs. -0.12), the $<$HR2$>$ values are essentially the same. There is a
trend that galaxies and clusters of galaxies seem to have somewhat
higher $<$HR2$>$ values than stars.

For $<$HR1$>$ more pronounced differences are found between the
individual classes of objects. Stars have on the average the lowest
HR1, i.e., they have the softest spectral energy distributions
(SED). This holds even, if one removes the white dwarfs and the
emission line stars (cf. Paper\,III) from the sample, since $<$HR1$>$
changes from +0.11 to +0.12 only. AGN display on the average 
($<$HR1$>$=+0.35) a somewhat harder SED than stars.

A significantly harder SED is found for clusters of galaxies with
$<$HR1$>$=+0.65$\pm$0.26. This is indeed what is expected for
this class of objects which exhibit a rather hard intrinsic spectrum
caused by thermal bremsstrahlung from a hot (10$^{7}$-10$^{8}$ K) plasma
(cf. e.g. B\"ohringer \cite{Boehringer96}).

For isolated galaxies $<$HR1$>$=+0.66$\pm$0.31 has been found. This is
nearly the same mean hardness ratio as found for clusters of galaxies
and that strongly supports the presumption discussed in Paper\,II that
most of the isolated galaxies found in the RASS are in fact members of
groups or even cluster of galaxies. The limited field of our optical
images may not always have allowed to recognize the cluster. The X-ray
data themselves provide little evidence, since the extension parameter
given by the ROSAT SASS is, as
discussed by Kneer (\cite{Kneer96}), of limited use only for the
identification of clusters of galaxies.

\begin{table*}
\caption[]{
Mean hardness ratios HR1 and HR2. 'No. of sources' gives the number of
sources which fulfill the selection criteria discussed in the text.}
\begin{tabular}{lrrrrrrr}
\noalign{\smallskip} \hline \noalign{\smallskip}
\multicolumn{8}{c}{study area}\\
&\multicolumn{1}{c}{I}&\multicolumn{1}{c}{II}&\multicolumn{1}{c}{III}
&\multicolumn{1}{c}{IV}&\multicolumn{1}{c}{V}&\multicolumn{1}{c}{VI}
&\multicolumn{1}{c}{total}\\
\noalign{\smallskip} \hline \noalign{\smallskip} 
\multicolumn{8}{c}{stars} \\
No. of sources 
&\multicolumn{1}{c}{52}&\multicolumn{1}{c}{44}&\multicolumn{1}{c}{36}
&\multicolumn{1}{c}{13}&\multicolumn{1}{c}{43}&\multicolumn{1}{c}{25}
&\multicolumn{1}{c}{213}\\
$<$HR1$>$ & +0.01$\pm$0.49 & +0.16$\pm$0.40 & +0.08$\pm$0.37 
& -0.09$\pm$0.25 & +0.14$\pm$0.37 & +0.30$\pm$0.53 & +0.11$\pm$0.43 \\
$<$HR2$>$ & +0.04$\pm$0.30 & +0.10$\pm$0.34 & +0.04$\pm$0.30
& +0.15$\pm$0.39 & -0.02$\pm$0.24 & +0.01$\pm$0.32 & +0.05$\pm$0.31 \\
\noalign{\smallskip} \hline \noalign{\smallskip} 
\multicolumn{8}{c}{AGN} \\
No. of sources 
&\multicolumn{1}{c}{13}&\multicolumn{1}{c}{40}&\multicolumn{1}{c}{47}
&\multicolumn{1}{c}{64}&\multicolumn{1}{c}{54}&\multicolumn{1}{c}{33}
&\multicolumn{1}{c}{251}\\
$<$HR1$>$ & +0.83$\pm$0.17 & +0.59$\pm$0.26 & +0.39$\pm$0.33 
& -0.13$\pm$0.34 & +0.43$\pm$0.34 & +0.65$\pm$0.27 & +0.35$\pm$0.43 \\
$<$HR2$>$ & +0.30$\pm$0.30 & +0.22$\pm$0.27 & +0.20$\pm$0.29
& +0.12$\pm$0.29 & +0.17$\pm$0.25 & +0.19$\pm$0.30 & +0.18$\pm$0.28 \\
\noalign{\smallskip} \hline
\end{tabular}
\label{idents}
\end{table*}

Taken both hardness ratios together,
there is no galaxy and only one cluster with 
negative values in both bands. Since this conclusion is based on a total
of 93 objects, one may safely conclude that two negative hardness
ratios are very rare for galaxies and cluster of galaxies, even in fields
with very low hydrogen column densities.

For the AGN pronounced variations of the $<$HR1$>$ values have
been found between the individual fields. For field IV
one gets a softer average X-ray spectrum of the AGN with
$<$HR1$>_{IV}$=-0.13$\pm$0.34, whereas for field I one gets 
a very hard AGN spectrum with $<$HR1$>_{I}$=+0.83$\pm$0.17 with
no HR1 below +0.39. For the whole sample individual HR1 values as low
as -0.83 were found. These differences of the HR1 hardness
ratios reflect mainly the differences in the absorbing
hydrogen column densities. For the other four fields the mean
ratios $<$HR1$>$ are between +0.39 and +0.65, i.e., in between the 
two extreme values. It is interesting that the second highest
value of +0.63 is found in field VI which exhibits the second
highest value of $N_{\rm H}$ too.

For the stars the situation is different. There are no
significant differences in the mean HR1 ratios between the individual
fields, not even between fields I and IV, for which significant
differences have been found for the AGN. The fact that there are no
differences in the mean hardness ratios between the stellar samples in
fields which differ by about a factor of seven in the hydrogen column
density wheras there are siginificant differences in the hardness
ratios for the extragalactic sources clearly shows that at least the
vast majority of all our stellar counterparts must be located in front
of the absorbing hydrogen. We note that the mean $N_{\rm H}$
values are not based on our X-ray data, but, as discussed in Paper
II, on radioastronomical measurements compiled by Dickey \& Lockman
(\cite{DickeyLockman90}).

With the exception of two sources there are no stellar sources in the
area of hardness ratios with both HR1 $\geq$ 0.0 and HR2 $\geq$ +0.5,
whereas the other three classes of objects fairly densely populate
this area.  With the exception of a few very soft sources there are
no stellar objects with HR1 $\leq$ -0.45. However, the very soft
sources are a somewhat special case, since they have the major
fraction of their X-ray flux in the soft band and only little or
nothing at all in the hard band. This means that usually HR2 is not
well defined and affected by a large error. If one includes also
sources with high errors in HR2 one obtains a total of 22 sources with
HR1 $\leq$ -0.5. Including these sources in the total sample of
stellar counterparts one obtains $<$HR1$>$=+0.03$\pm0.47$ which is
slightly lower than the value given in Table 4.

\begin{table}[ht]
\caption[]{
Mean hardness ratios HR1 and HR2 for galaxies, clusters and 
X-ray sources without optical counterparts for the total sample. }
\begin{tabular}{lrrr}
\noalign{\smallskip} \hline \noalign{\smallskip}
\noalign{\smallskip} \hline \noalign{\smallskip} 
&\multicolumn{1}{c}{galaxies}&\multicolumn{1}{c}{clusters}
&\multicolumn{1}{c}{no ident.}   \\
No. of sources: & \multicolumn{1}{c}{25}& \multicolumn{1}{c}{72}
& \multicolumn{1}{c}{9} \\
$<$HR1$>$ & +0.66$\pm$0.31 & +0.65$\pm$0.26 & +0.43$\pm$0.39  \\
$<$HR2$>$ & +0.27$\pm$0.39 & +0.22$\pm$0.30 & +0.26$\pm$0.39  \\
\noalign{\smallskip} \hline
\end{tabular}
\label{idents}
\end{table}

\subsection{X-ray-to-optical flux ratios}
\label{X-ray-to-optical flux ratios}

As already noted by earlier studies (e.g. Maccacaro et
al. \cite{Maccacaroetal88} or Stocke et al. \cite{Stockeetal91}) the
X-ray-to-optical flux ratios differ significantly between the object
classes. Therefore, for the identification of their objects they used
this criterion in order to pre-select the sources they actually
observed. While this methods leads in many cases to reliable
pre-identification, it is nevertheless obvious that this pre-selection
does also bias any statistical derivation of the X-ray-to-optical flux
ratio.

In the present study, as discussed in Section 4 of Paper\,II, a
somewhat different approach had been chosen for the identification of
the optical counterparts. There was no pre-selection according to the
X-ray-to-optical flux ratio. Only in doubtful cases the
X-ray-to-optical flux ratio was used (as well as hardness ratios HR1
and HR2) as secondary criterion for the identification. Hence, the
X-ray-to-optical flux ratios of our sample should be less biased
than those of the EMSS.

\begin{figure*}
\psfig{figure=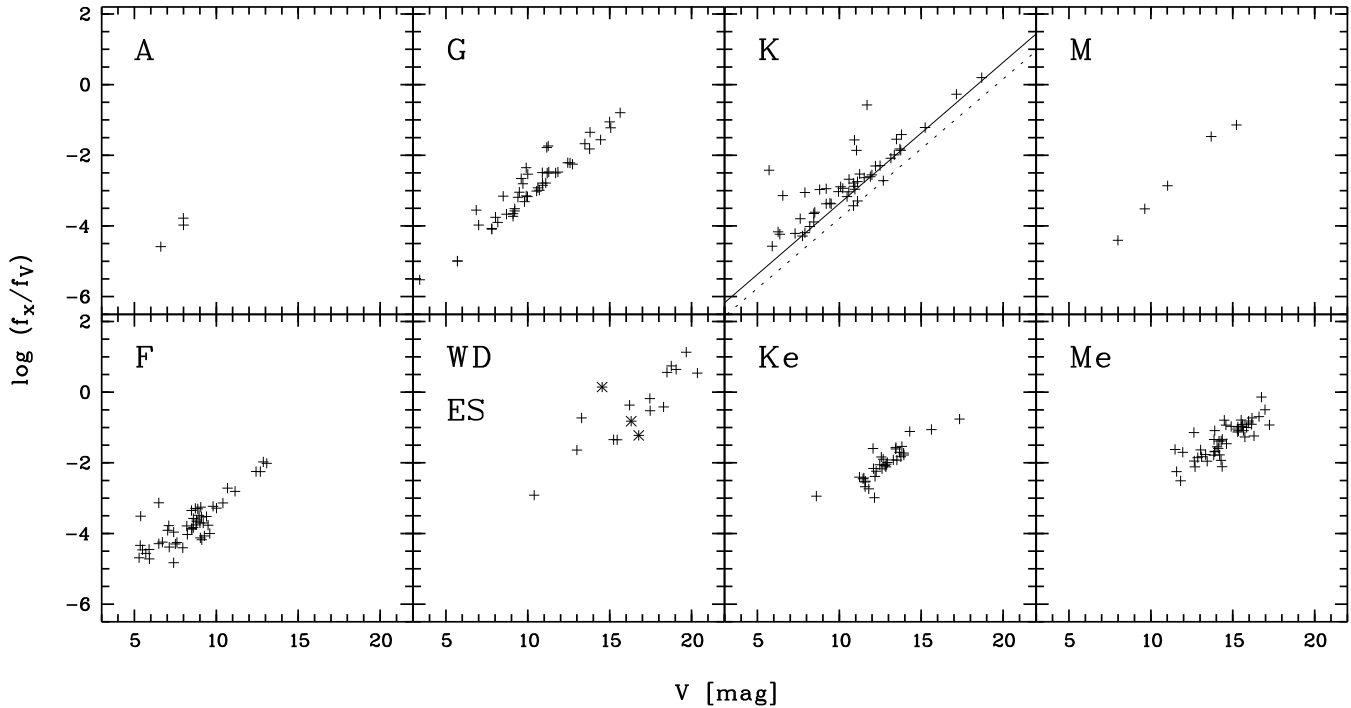,width=18.0cm,angle=-90,bbllx=136pt,bblly=28pt,bburx=534pt,bbury=770pt,clip=}
\caption[]{X-ray-to-optical flux ratios 
log$[f_{x}$/$f_{V}]$
vs. the optical flux $f_{V}$ for the stars in our sample. The letters 
denote the spectral type. 'WD' means white dwarf (asterisks) and 'ES' emission
line star (crosses). The lines plotted in the diagram of the K stars are
lower flux limit for coronal sources (see text). 
}
\label{jk8432.f2}
\end{figure*}

\begin{figure}
\psfig{figure=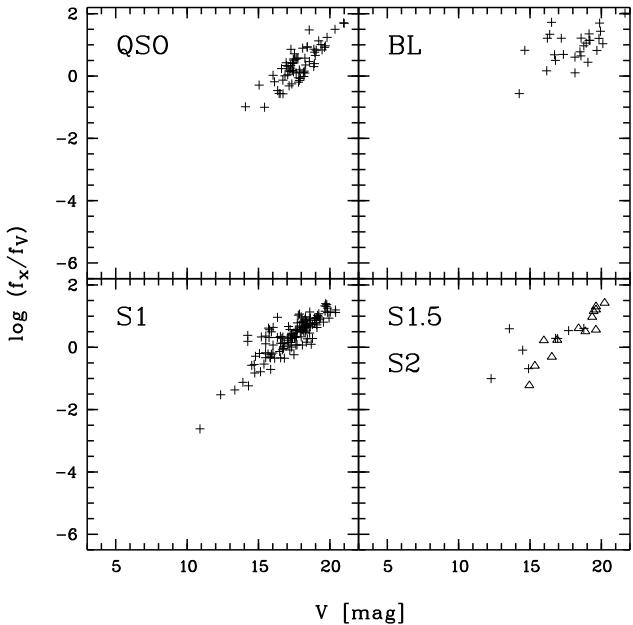,width=8.5cm,angle=-90,bbllx=136pt,bblly=28pt,bburx=534pt,bbury=430pt,clip=}
\caption[]{X-ray-to-optical flux ratios 
log$[f_{x}$/$f_{V}]$
vs. the optical flux $f_{V}$ for various classes of AGN. 
}
\label{jk8432.f3}
\end{figure}

The conversion of count rates to fluxes which depends on the intrinsic
spectral energy distribution and on the hydrogen column density has
been extensively discussed in Paper\,II.
Figs. \ref{jk8432.f2} and \ref{jk8432.f3} show X-ray-to-optical 
flux ratios 
log$[f_{x}$/$f_{V}]$
vs. the optical flux $f_{V}$ for various object classes. In Table 6
the limits of the log[$f_{x}$/$f_{V}$] ratios as well as average
values are given. Clusters of galaxies are not included, since,
contrary to the EMSS sample, in our case usually no brightness has
been determined for the individual members of the clusters.

The lower boundaries of the distributions shown in Figs. \ref{jk8432.f2} 
and \ref{jk8432.f3} are
determined by the count limit chosen. As already mentioned, they
depend on the spectral energy distribution assumed and the hydrogen
column density. For demonstration purposes we have plotted in the
diagram of the K stars the lower limit for coronal sources with the
flux limit of 1.8$\cdot$10$^{-13}$\,erg\,s$^{-1}$\,cm$^{-2}$ (solid
line) given in Table 3 of Paper\,II. In addition, we have also plotted
the lower limit for 0.6$\cdot$10$^{-13}$\,erg\,s$^{-1}$\,cm$^{-2}$ for
field V (dashed line) which has the lowest hydrogen column density.
It is obvious that the sources below the lower boundaries are missed
in our sample because of the count rate limit.

\begin{table}[ht]
\caption[]{Ranges and average values of X-ray-to-optical flux ratios for
various classes of objects.} 
\begin{tabular}{lrrr} 
\noalign{\smallskip} \hline \noalign{\smallskip}
Object class & \multicolumn{2}{c}{log$[f_{x}$/$f_{V}]_{limits}$}
& \multicolumn{1}{c}{$<log[f_{x}$/$f_{V}]>$} \\
& \multicolumn{1}{c}{min} & \multicolumn{1}{c}{max} \\ 
\noalign{\smallskip} \hline \noalign{\smallskip} 
stars$_{tot}$ & -5.52 & +1.13 & -2.46$\pm$1.27 \\
A  stars      & -4.59 & -3.78 & -4.11$\pm$0.42 \\
F  stars      & -4.82 & -1.97 & -3.70$\pm$0.68 \\
G  stars      & -5.52 & -0.80 & -2.89$\pm$1.02 \\
K  stars      & -4.57 & +0.20 & -2.77$\pm$1.04 \\
M  stars      & -4.41 & -1.14 & -2.68$\pm$1.37 \\
Ke stars      & -2.99 & -0.76 & -2.02$\pm$0.51 \\
Me stars      & -2.51 & -0.14 & -1.33$\pm$0.50 \\
WD            & -1.22 & +0.14 & -0.63$\pm$0.70 \\
ES            & -2.91 & +1.13 & -0.42$\pm$1.12 \\
AGN$_{tot}$   & -2.62 & +2.02 & +0.41$\pm$0.65 \\
S1            & -2.62 & +1.40 & +0.37$\pm$0.61 \\
S1.5          & -1.01 & +0.61 & +0.07$\pm$0.61 \\
S2            & -1.22 & +1.43 & +0.47$\pm$0.86 \\
QSO           & -1.00 & +1.71 & +0.35$\pm$0.57 \\
BL Lac        & -0.56 & +2.02 & +0.94$\pm$0.54 \\
Galaxies      & -2.14 & +1.61 & -0.27$\pm$0.92 \\
\noalign{\smallskip} \hline
\end{tabular}
\label{idents}
\end{table}

\begin{figure}
\psfig{figure=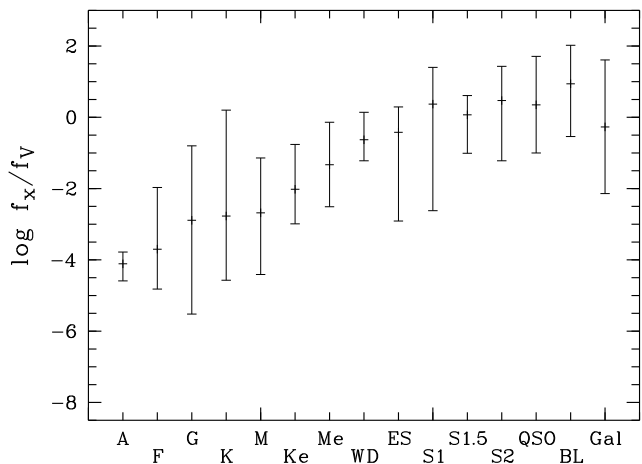,width=8.5cm,angle=-90,bbllx=137pt,bblly=44pt,bburx=520pt,bbury=572pt,clip=}
\caption[]{Mean values and ranges of $\log f_{\rm x}/f_V$ as listed in 
Table 6 for the different classes of X-ray emitters. The various 
classes overlap considerably. 
}
\label{jk8432.f4}
\end{figure}

As compared with the EMSS, most object classes in the RASS sample
display a somewhat wider range in the $<log[f_{x}$/$f_{V}]>$ ratios,
even if one takes into account that in the numbers given in Stocke et
al.'s Table 1 objects with extreme values are excluded. This means
that in the RASS samples there are larger overlapping areas which do
not allow an unequivocal pre-selection on the basis of the
$<log[f_{x}$/$f_{V}]>$ ratios. 
This is shown in Fig. \ref{jk8432.f4} which displays the ranges of 
$\log f_{\rm x}/f_V$ listed in Table 6.

In particular we would like to note
that there is a large overlapping area between the stars and the AGN,
even if one excludes white dwarfs and cataclysmic binaries which show
the highest $<log[f_{x}$/$f_{V}]>$ ratios among the stellar
counterparts.

The large difference for the maximal values for galaxies can be
explained by the fact that a significant fraction of the galaxies in
the RASS sample are, as discussed above, indeed groups or clusters of
galaxies.  Another striking difference is the presence of a
significant number of AGN with $<log[f_{x}$/$f_{V}]> \>$ 1.  In the
EMSS only very few AGN with such high ratios of X-ray-to-optical flux
have been found. A possible explanation for this difference could be
that the limiting magnitudes of the optical identifications for the
RASS sample are lower than those of the EMSS, since the highest
X-ray-to-optical flux ratios are indeed found for the faintest
objects.

As one would expect for the stellar
counterparts in our sample (cf. e.g. Schmitt \cite{Schmitt90}), 
the maximal value of the
$<log[f_{x}$/$f_{V}]>$ ratio increases with decreasing T$_{eff}$. On
the average, Ke and Me stars show, as expected (see 2.2), 
higher X-ray luminosities than the K
and M stars without emission lines. The Ke and Me stars display a 
much smaller
range of $<log[f_{x}$/$f_{V}]>$ ratios than the normal K and M stars
which is either due to the fact that no X-ray faint Ke and Me 
stars are found or simply, that there were no optically bright emission
line stars in our sample.

\subsection{Stellar counterparts }
\label{Stars}

\begin{figure}
\psfig{figure=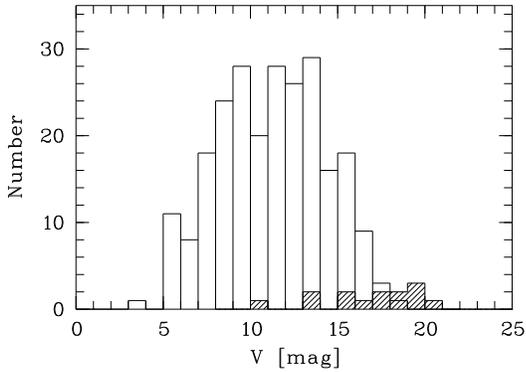,width=8.0cm,angle=-90,bbllx=187pt,bblly=34pt,bburx=538pt}
\caption[]{Histogram showing the number distribution of the visual
brightness of the stellar counterparts. The class "ES" is separately
plotted as hashed histogram.
}
\label{jk8432.f5}
\end{figure}

Fig. \ref{jk8432.f5} shows the visual brightness distribution of the
stellar counterparts. It peaks around around 11th to 13th magnitude
which is due to the large number of K and M stars in this brightness
range. Among the faintest counterparts ($V \ga 17$) objects classified
"ES" dominate, i.e. mostly cataclysmic variables (hashed histogram in
Fig. \ref{jk8432.f5}).

\begin{figure}
\psfig{figure=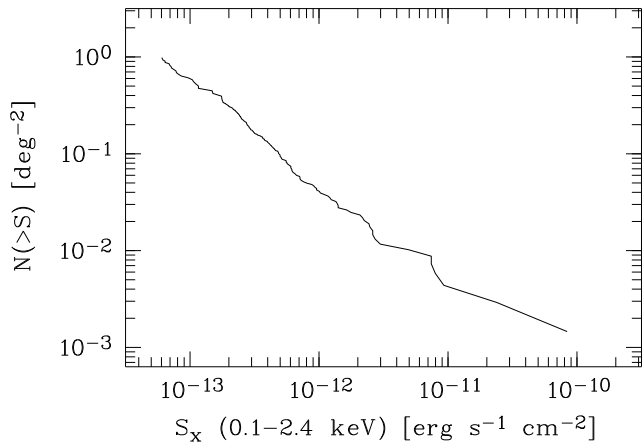,width=8.5cm,angle=-90,bbllx=50pt,bblly=65pt,bburx=540pt,bbury=770pt,clip=}
\caption[]{Area-corrected $\log\,N\,-\,\log\,S$ distribution for stellar
counterparts (exluding classes ``ES'' and ``WD'').}
\label{jk8432.f6}
\end{figure}

Fig. \ref{jk8432.f6} shows the $\log\,N\,-\,\log\,S$ distribution for
the stellar sample, excluding objects of classes ``ES'' and ``WD''.
For fluxes below $1.8\,10^{-13}$\,erg\,s$^{-1}$\,cm$^{-2}$ the smaller
survey area of study area V has been taken into account (cf. Paper
II). The slope in the range $S_{\rm x} = 2\,10^{-13} -
10^{-11}$\,erg\,s$^{-1}$\,cm$^{-2}$ is $-1.36\pm0.11$ (maximum
likelihood), hence slightly flatter than expected for the Euclidian
slope of $-1.5$. This indicates that our sample is affected by the
scale height of the galactic distribution of the stellar counterparts.

\subsection{Active Galactic Nuclei}
\label{AGN}
\subsubsection{Redshift distribution}

\begin{figure}
\psfig{figure=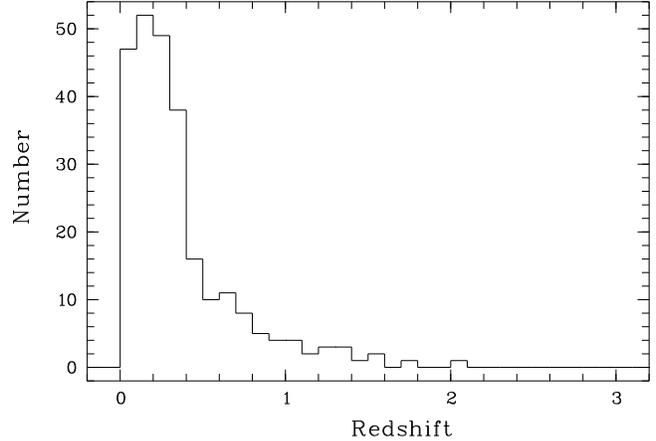,width=8.5cm,angle=-90,bbllx=50pt,bblly=30pt,bburx=546pt,bbury=742pt,clip=}
\caption[]{Redshift distribution of the AGN of the whole sample. Not included
is the quasar at z=4.28
}
\label{jk8432.f7}
\end{figure}

The redshift distribution of the AGN of the whole sample is presented
in Fig. \ref{jk8432.f7}.  As can be seen, the majority of the AGN in our sample
are found at rather low redshifts with the median value of the whole
sample being z$_{med}$=0.24. The median values for the individual
fields vary between z$_{med}$=0.15 and z$_{med}$=0.36. The highest
value of z$_{med}$=0.36 is found in field V where the minimum
count rate of 0.01 cts\,s$^{-1}$ is lower by a factor of 3 as compared
to 0.03 cts\,s$^{-1}$ in the other five fields. The second highest
median z value z$_{med}$=0.31 is found in field IV which is the field
with the lowest $N_{\rm H}$. Correspondingly, the lowest median value
z$_{med}$=0.15 for the redshift distribution is found in field I, the
field with the highest $N_{\rm H}$.

We note that the highest redshift z=4.28 found for the quasar
RXJ1028.6-0844 (Zickgraf et al.  \cite{Zickgrafetal97b}) is one of the
highest redshifts at all found for RASS sources.

\subsubsection{$\log\,N\,-\,\log\,S$ distributions}
Because of the limited RASS exposure time as well as the flux limit
chosen by us (cf. Paper\,II), only the brighter AGN show up in our
RASS sample. Hence, our sample is the ideal complement to the deep
surveys where preferentially low-luminosity AGN are found. Both data
sets combined allow to construct a $\log\,N\,-\,\log\,S$ function
which covers a wide range of luminosities and which can be used for
cosmological studies.  This has been done and discussed by Lehmann et
al. \cite{Lehmannetal98} and Miyaji et al. \cite{Miyajietal98} who
combined our sample with 207 ksec PSPC observations of the Lockman
Hole and with the 1.112 Msec HRI ultradeep survey observations
(Hasinger et al. \cite{Hasingeretal98}).



%
%

\begin{figure}
\psfig{figure=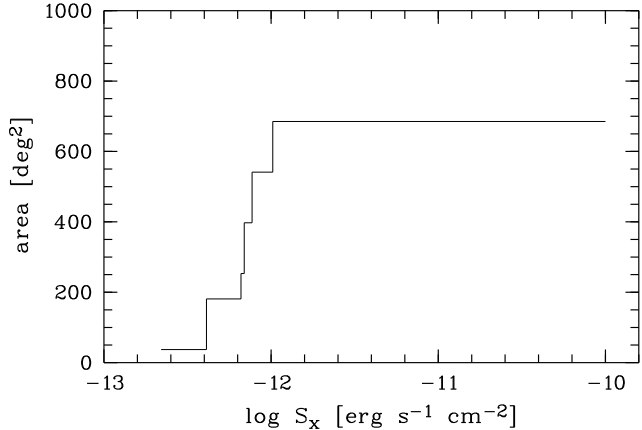,width=8.5cm,angle=-90,bbllx=187pt,bblly=34pt,bburx=538pt,bbury=544pt,clip=}
\caption[]{Survey area as a function of flux. The limiting fluxes for the
individual study areas were taken from Paper\,II. 
}
\label{jk8432.f8}
\end{figure}

\begin{figure}
\psfig{figure=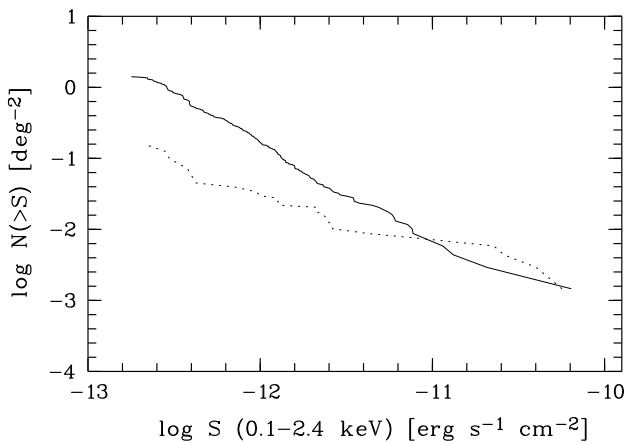,width=8.5cm,angle=-90,bbllx=187pt,bblly=33pt,bburx=540pt,bbury=544pt,clip=}
\caption[]{Area-corrected $\log\,N\,-\,\log\,S$ distributions for AGN
 (solid line) and BL Lacs (dashed line). 
}
\label{jk8432.f9}
\end{figure}

In order to correct for the variation of the survey area as a function
of flux, we computed $\log\,N\,-\,\log\,S$ distributions by taking the
area as displayed in Fig. \ref{jk8432.f8} into account. We used the
flux limits given in Paper\,II for each of the individual study
areas. The area drops from the total of 685\,deg$^2$ at fluxes
10$^{-12}$ \ergscm\ in several steps at the individual limiting fluxes
to 37\,deg$^2$, which is the size of area V. The area corrected 
$\log\,N\,-\,\log\,S$ distributions for all AGN and for the BL Lac objects
in our sample are diplayed in Fig. \ref{jk8432.f9}. The
distributions are linearly increasing towards the lowest flux. For
the AGN no indication of a flattening due to incompleteness is
indicated except at the lowest fluxes.

The slope of the $\log\,N\,-\,\log\,S$ distributions for AGN is with
$-1.44 \pm0.09$ in good agreement with the euclidian slope of $-1.5$.
The BL Lacs exhibit with a slope of $-0.72\pm0.13$ a much flatter
distribution. The errors have been calculated using the 
formula given
by Crawford et al. (\cite{Crawfordetal70}) basing on the maximum-likelihood 
technique. 
We suspect that the much lower slope of the BL Lac objects
is due to incompleteness.  In
addition, it could be that a major fraction of
the 12 undidentified sources are in fact BL Lac objects.  Moreover, due
to the low number (30) of BL Lacs our $\log\,N\,-\,\log\,S$
distribution is affected by statistical uncertainties. 

A rather flat $\log\,N\,-\,\log\,S$
distribution for BL Lacs has been also found by
Bade et al. \cite{Badeetal98}, who studied a sample of 37 BL Lac
objects from the RASS. In their sample the $\log\,N\,-\,\log\,S$ 
curve starts
to flatten at 8$\cdot$10$^{-12}$\,erg\,cm$^{-2}$\,s$^{-1}$ which,
taking into account their different energy range (0.5-2.0 keV) 
this
corresponds to 1.8 $\cdot$10$^{-12}$\,erg\,cm$^{-2}$\,s$^{-1}$ in 
our energy range (0.07-2.4 keV). However,
contrary to the above conclusions for our sample they do not ascribe
this flattening to incompleteness. For our RASS sample the
$\log\,N\,-\,\log\,S$
curve for BL Lacs lies slightly above the curve
found by Bade et al.  Unfortunately, it is not possible to compare the
redshift and luminosity distributions of the two samples, since the
low resolution spectra used for our optical identification did not
allow us to derive the redshifts of the BL Lac objects.

\subsubsection{Radio fluxes} 
\begin{figure}
\psfig{figure=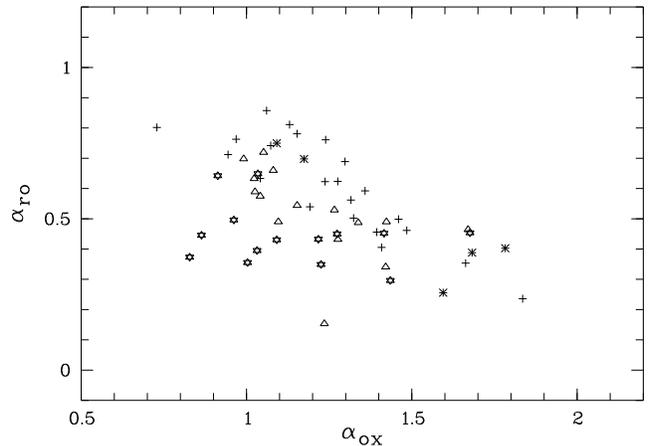,width=8.5cm,angle=-90,bbllx=50pt,bblly=34pt,bburx=535pt,bbury=720pt,clip=}
\caption[]{$\alpha_{\rm ro}$-$\alpha_{\rm ox}$ diagram  for QSOs ($+$), Seyfert
galaxies (triangles), BL Lacs (star symbols), and radio galaxies (asterisks).
}
\label{jk8432.f10}
\end{figure}

The radio fluxes at 4.85\,GHz, the X-ray flux $f_{\rm x}$ in the
0.1-2.4\,keV energy range, and the $B$ magnitudes given in Paper\,III  
were used to calculate the continuum slopes $\alpha_{\rm ro}$ and  
$\alpha_{\rm ox}$. The definition of these coefficients is given e.g. in 
Stocke et al. (\cite{Stockeetal91}).   
The optical fluxes at 2500\,\AA\ were derived from the the $B$ magnitudes using
the relation given by Schmidt (\cite{Schmidt68}) for $\alpha_{\rm o} = -0.5$. 
For BL Lacs we assumed 
$\alpha_{\rm o} = -1.4$ (Ghisellini et al. \cite{Ghisellinietal86}).
The monochromatic X-ray flux at 2\,keV was obtained from the integrated flux
$f_{\rm x}$ using a photon index of -2. Radio fluxes at 4.85\,GHz from 
the 87GB catalog
listed in Paper III were K-corrected for spectral index $\alpha_{\rm
r} = -0.5$ for QSOs and Seyfert galaxies, 0.0 for BL Lacs, and -0.7
for galaxies.  For BL Lacs a redshift of 0.3 was assumed for those
without measured $z$.  The resulting $\alpha_{\rm ro}$-$\alpha_{\rm
ox}$ diagram is shown in Fig.  \ref{jk8432.f10}.

The sources are distributed in a similar way as shown in Stocke et al.
(\cite{Stockeetal91})
(cf. their Fig. 6) with the exception that our sample contains only a 
few radio-quiet objects (with respect to the limits shown in Stocke et
al.).
This is due to the flux limit of the 87GB catalog. Most of the BL Lacs
are 
located in a region between $\alpha_{\rm ro} = 0.3$ and 0.5 which is 
separated from that occupied by the radio-loud QSOs and Seyferts,
although 
some overlap exists in particular at large $\alpha_{\rm ox}$. The
extension 
of the BL Lac region towards larger $\alpha_{\rm ox}$ also contains
three 
radio galaxies. At least for the X-ray bright BL Lacs 
(i.e. small $\alpha_{\rm ox}$) a reliable classification based on the 
$\alpha_{\rm ox}$ - $\alpha_{\rm ro}$ diagram is possible. Likewise, both 
radio-loud and X-ray-bright AGN are well separated in the upper left
corner 
of the diagram.

\section{Conclusions}
\label{Conclusions}
The results of a complete optical identification of a spatially
complete, count rate limited of ROSAT All-sky survey X-ray sources
show that the RASS is dominated by stellar sources and AGN. While
nearly the same fraction of stars and AGN (40.7\% and 42.1\%,
respectively) are found in our total sample which comprises six
individual fields outside the galactic plane, distinct variations are
found between the individual fields. The main parameter responsible
for these variations is the mean hydrogen column density $N_{\rm H}$ in
the field which, however affects more the number of AGN than the
stars.  As the distribution of the mean hardness ratios $<$HR1$>$
shows, at least the vast majority of the stellar sources must be
located in front of the absorbing hydrogen.

For the accuracy of the RASS positions, i.e., the total distance
between optical source and X-ray position, a 90\% error circle of
about 30\arcsec\ has been found.  This error circle is considerably
lower than the 40-45\arcsec\ found by the SASS which is basing on
X-ray data alone.


Hardness and X-ray-to-optical flux ratios vary 
between the different object classes. However, since there are in the
corresponding parameter spaces large overlapping areas between the
different classes, an unambiguous identification on the basis of the
X-ray data alone is not possible.  Optical follow-up observations are
therefore indispensable for an identification.

RASS data are the ideal complement to deep surveys for
e.g. constructing $\log\,N\,-\,\log\,S$ distributions for cosmological
studies. In the RASS mainly objects at the bright end of the X-ray
luminosity function are found. However, because of its large spatial
area a significant number of the high luminosity objects which have a
low space density are found.

Because of the low resolution used for the spectroscopy in this
identification project more detailed investigations, such as searches
for Li $\lambda$6707 absorption in stellar sources were not
possible. However, as first publications based on our catalog show,
the results of our identification program offers a rich mine for many
more detailed follow-up studies.

\acknowledgements{
We would like to thank the observers who contributed to
the optical observations: J.M. Alcala, C. Alvarez, H. Bock, H. Bravo,  
O. Cardona, M. Coayahuitl, L. Corral, L. de la Cruz, U. Erkens, 
C. Fendt, A. Flores, A. Gallegos, T. G\"ang, J. Guichard, J. Heidt, 
M. K\"ummel, R. Madejski, 
A. Marquez, O. Martinez,  
A. Piceno, A. Porras, Th. Szeifert, J. R. Valdes,  F. Valera, G. 
Vazquez, R. Wichmann, K. Wilke, and O. Yam. We also would like to thank 
the staff of 
the Guillermo Haro Observatory for their friendly and helpful
support during the observations.
We thank H.T. MacGillivray at ROE, R.G. Cruddace and D.J. Yentis at NRL,
and R. McMahon at IoA for making available the COSMOS UKST
and APM data, respectively. This research also made use of the NASA/IPAC
Extragalactic Database (NED) which is operated by the Jet Propulsion
Laboratory, California Institute of Technology, under contract with NASA.
This work was supported by DARA under grant Verbundforschung 50\,OR\,90\,017. 
}


\begin{thebibliography}{}


\bibitem[1998]{Appenzelleretal98} Appenzeller, I., Thiering, I.,
Zickgraf, F.-J., et al., 1998, ApJS 117, 319 (Paper\,III)

\bibitem[1988]{Aschenbach88} Aschenbach, B., 1988, Appl.\,Optics, 27,
1404

\bibitem[1998]{Badeetal98} Bade, N., Beckmann, V., Douglas, N.C., Barthel,
P.D., Engels, D., Cordis, L., Nass, P., Voges, W., 1998, A\&A 334,
459

\bibitem[1996]{Boehringer96} B\"ohringer, H., 1996.
In: H.U. Zimmermann, J.E. Tr\"umper, H. Yorke (eds.), MPE
Rreport 263, ``R\"ontgenstrahlung from the Universe'', 
Garching 1996, p. 537

\bibitem[1970]{Crawfordetal70} Crawford, D.F., Jauncey, D.L., 
Murdoch, H.S., 1970, ApJ 162, 405

\bibitem[1990]{DickeyLockman90} Dickey J.M., Lockman F.J., 1990,
ARA{\&}A 28, 215

\bibitem[1986]{Ghisellinietal86}
Ghisellini G., Maraschi L., Tanzi E.G., Treves A., 1986, ApJ 310, 317

\bibitem[1996]{Haisch96} Haisch, B., Schmitt, J.H.M.M., 1996,
PASP 108, 113


\bibitem[1998]{Hasingeretal98} Hasinger G.R., Burg R., Giacconi R., et
al., 1998, A\&A 329, 482 

\bibitem[1996]{Kneer96} Kneer R., 1996, Ph.D. thesis, University of
Heidelberg

\bibitem[1998]{Lehmannetal98} Lehmann, I., Hasinger, G., Schwope, A.,
Boller, T., 1998, in: Proceedings of the Symposium
``Highlights in X-ray Astronomy in honour of
Joachim Tr\"umper's 65th birthday'',
eds. B. Aschenbach \& M.J. Freyberg, MPE Report 269 (in press)

\bibitem[1988]{Maccacaroetal88} Maccacaro, T., Gioia, I., Wolter, A., 
Zamorani, G., Stocke, J., 1988, ApJ 326, 680

\bibitem[1091]{MihalasBinney81}Mihalas, D., Binney, J., Galactic
Astronomy, 1981, W.\,H.\,Freeman and Company, San Francisco 

\bibitem[1998]{Miyajietal98}Miyaji, T., Hasinger, G., Schmidt, M.,
1998, in: Proceedings of the Symposium
``Highlights in X-ray Astronomy in honour of
Joachim Tr\"umper's 65th birthday'',
eds. B. Aschenbach \& M.J. Freyberg, MPE Report 269 (in press) 

\bibitem[1968]{Schmidt68}Schmidt, M., 1968, ApJ 151, 393

\bibitem[1990]{Schmitt90} Schmitt, J.H.M.M., 1990, Adv. Space Res. 10, 
115

\bibitem[1991]{Stockeetal91}
Stocke, J.T., Morris, S.L., Gioia, I.M., et al., 1991, ApJS 76, 813

\bibitem[1983]{Trumper83} 
Tr\"umper, J., 1983, Adv.\,Space\,Res. vol. 2, No.\,4, 241

\bibitem[1999]{Vogesetal99} 
Voges W., Aschenbach B., Boller, Th., et al., 1999 A\&A 349, 389

\bibitem[1997]{Zickgrafetal97a}
Zickgraf, F.-J., Thiering, I., Krautter, J., et al., 1997, A\&AS 123,
103 (Paper\,II)

\bibitem[1997]{Zickgrafetal97b}
Zickgraf, F.-J., Voges, W., Krautter, J., Thiering, I.,
Appenzeller, I., Mujica, R., Serrano, A., 1997, A\&A 323, L21

\bibitem[1998]{Zickgrafetal98}
Zickgraf, F.-J., Alcala\'{A}, J.M., Krautter, J., Appenzeller, I., 
Sterzik, M.F., Motch, C., Pakull, M.W. 1998, A\&A 339, 457 

\end{thebibliography}
\end{document}